%% file: Main.tex
\documentclass[conference]{IEEEtran}
\IEEEoverridecommandlockouts

\usepackage{cite}
\usepackage{amsmath,amssymb,amsfonts}
\usepackage{pifont}
\usepackage{algorithmic}
\usepackage{graphicx}
\usepackage{textcomp}
\usepackage{xcolor}
\usepackage{url}
\usepackage{booktabs}
\usepackage{multirow}
\usepackage{makecell}
\usepackage{listings}
\usepackage[T1]{fontenc}
\usepackage[scaled=0.9]{beramono}
\makeatletter
\DeclareUrlCommand\code{\urlstyle{tt}}
\g@addto@macro\UrlBreaks{\do\.\do\:\do\_\do\,\do\(\do\)}
\makeatother
\def\BibTeX{{\rm B\kern-.05em{\sc i\kern-.025em b}\kern-.08em
    T\kern-.1667em\lower.7ex\hbox{E}\kern-.125emX}}

\begin{document}

\title{AsyncSparse: Accelerating Sparse Matrix-Matrix Multiplication on Asynchronous GPU Architectures}

\author{
    \IEEEauthorblockN{Jie Liu\IEEEauthorrefmark{1}, Huanzhi Pu\IEEEauthorrefmark{2}, Zhiru Zhang\IEEEauthorrefmark{1}}
    \IEEEauthorblockA{\IEEEauthorrefmark{1}Cornell University \quad
      \IEEEauthorrefmark{2}Georgia Institute of Technology}
  }

\maketitle

\input{Sec-Abstract}


\input{Sec1-Intro}
\input{Sec2-Background}
\input{Sec3-Method}
\input{Sec4-Evaluation}

\input{Sec5-Limitations}

\input{Sec6-Conclusion}

\section*{Acknowledgment}
An AI assistant (Claude, Anthropic) was used to help improve the writing in Sections~\ref{sec:introduction}, \ref{sec:background}, \ref{sec:method}, \ref{sec:evaluation}, \ref{sec:related} and \ref{sec:conclusion}.

\bibliographystyle{IEEEtran}
\bibliography{ref}

\end{document}

%% file: Sec-Abstract.tex
\begin{abstract}
Sparse Matrix-Matrix Multiplication (SpMM) is a fundamental kernel across scientific computing and machine learning. While prior work accelerates SpMM using Tensor Cores, no existing sparse kernel exploits the asynchronous features of modern GPU architectures, such as NVIDIA's Tensor Memory Accelerator (TMA) and warp specialization. This work systematically studies how these features impact SpMM performance and introduces two co-designed kernels. For structured sparsity, we optimize a warp-specialized producer-consumer pipeline overlapping TMA data transfer with WGMMA computation using Block Compressed Sparse Row (BCSR) format. For irregular sparsity, we design a Window Compressed Sparse Row (WCSR) kernel that loads the sparse operand via TMA and splits large row-windows across thread blocks for load balancing. Our WCSR kernel outperforms all prior SpMM kernels on SuiteSparse matrices (1.47$\times$ over AccSpMM, 6.24$\times$ over cuSPARSE). Our BCSR kernel achieves a combined 2.66$\times$ end-to-end speedup on Qwen2.5-7B prefill at 90\% block sparsity with 64K tokens over cuDNN/cuBLAS.
\end{abstract}

%% file: Sec1-Intro.tex
\section{Introduction}\label{sec:introduction}
Sparse Matrix-Matrix Multiplication (SpMM) is a performance-critical primitive across scientific computing~\cite{saad2003iterative,duff2002overview}, graph analytics~\cite{kepner2016mathematical,wang2016gunrock}, and deep learning inference~\cite{hoefler2021sparsity,gale2020sparse,wang2019deep}.
As problem sizes and model scales grow, SpMM performance on GPU accelerators increasingly dominates end-to-end runtime and cost.
To harness the massive computate power of contemporary GPUs, applications frequently enforce block sparsity, where nonzero elements in the sparse operand are clustered into dense sub-blocks~\cite{narang2017block,gray2017gpu,jeong2025enabling}.
In deep learning, structured pruning techniques~\cite{frantar2023sparsegpt,sun2023simple} drive this trend, reducing the inference cost while maintaining accuracy.
This structural regularity bridges the gap between memory-bound sparse operations and compute-bound dense General Matrix-Matrix Multiplication (GEMM) by enabling SpMM to be realized as a sequence of micro-GEMMs.

However, realizing this theoretical advantage on modern GPUs is challenging.
Recent work in high-performance computing increasingly relies on specialized GPU features (e.g., Tensor Cores) to accelerate sparse kernels~\cite{tcgnn, shi2025flashsparse, zhao2025acc, jeong2025enabling}, yet prior approaches largely overlook the key trend in massively parallel architectures: specialization and asynchronous execution.
The NVIDIA Hopper architecture~\cite{luo2025dissecting} exemplifies this shift with the Tensor Memory Accelerator (TMA) for asynchronous bulk data movement and Warpgroup Matrix-Multiply-Accumulate (WGMMA) for high-throughput Tensor Core execution.
Effectively mapping block sparsity onto this asynchronous pipeline requires careful overlapping of data movement and computation, coordinated barrier-based synchronization, and efficient tensor layouts across the GPU memory hierarchy.

Existing production libraries and research kernels do not fully exploit this model.
cuSPARSE~\cite{naumov2010cusparse} primarily targets the Blocked-ELLPACK (BELL) format, which enforces uniform block-row lengths by padding with explicit zero blocks; on irregular sparsity distributions this leads to substantial compute waste.
PyTorch's TorchAO~\cite{or2025torchao} supports Block Compressed Sparse Row (BCSR) without padding, but its Triton-generated kernel underutilizes Hopper's asynchronous pipelines and falls short of saturating WGMMA.
Recent research kernels exploits Tensor Cores on modern GPUs~\cite{jeong2025enabling, zhao2025acc, xia2023flash, fan2024dtc, shi2025flashsparse}, but none adopt the asynchronous programming paradigm, leaving significant performance gap on the table.

In this paper, we present AsyncSparse, a set of high-performance SpMM kernels co-designed with the asynchronous execution model of the NVIDIA Hopper architecture.
We use two complementary sparse formats, BCSR and Window Compressed Sparse Row (WCSR), as vehicles to analyze the performance implications of asynchronous GPU features on SpMM.
For the BCSR kernel, both operands reside in contiguous global memory, enabling a warp-specialized producer--consumer pipeline where the producer warpgroup asynchronously gathers both operands via TMA while two consumer warpgroups execute back-to-back WGMMA instructions over a multi-stage circular buffer in shared memory.
For WCSR, a finer-grained format that compresses column vectors per row window, the dense operand requires indirect address offset logic that TMA cannot perform.
We therefore implement a single warp-group design where all threads cooperatively load the dense operand, and thread blocks compute split chunks across windows for better load balance.

Beyond delivering high-performance kernels, we conduct a systematic performance study that dissects the impact of each Hopper architectural feature on SpMM throughput.
To our knowledge, this is the first work to provide a comprehensive performance study for SpMM on asynchronous GPU programming models.

Specifically, this paper makes the following contributions:
\begin{itemize}
\item \textbf{Warp-specialized BCSR SpMM kernel.}
We design a producer--consumer warp-specialized pipeline for BCSR SpMM that overlaps TMA-driven memory gathering with WGMMA Tensor Core computation via a multi-stage asynchronous circular buffer, with significantly less zero-padding overhead than the BELL format while sustaining high Tensor Core utilization.
\item \textbf{Load-balanced WCSR SpMM kernel.}
We design a WCSR SpMM kernel that loads the sparse operand via TMA and uses cooperative thread gathering for the dense operand, with task-based decomposition that splits large row-windows into fixed-size sub-tasks to balance irregular workloads across thread blocks.
\item \textbf{Systematic architectural study.}
We analyze the performance impact of modern asynchronous GPU features on SpMM throughput. A detailed ablation study on our BF16 SpMM kernel shows that WGMMA, TMA, and warp specialization collectively account for $\approx 98\%$ of the total performance gain over a standard CUDA-core baseline.
\item \textbf{State-of-the-art performance.}
We evaluate our kernels on an NVIDIA H100 GPU across 414 SuiteSparse matrices, where our WCSR kernel achieves a geometric-mean speedup of $1.47{\times}$ over AccSpMM and $6.24{\times}$ over cuSPARSE at dense-matrix width $N{=}1024$. Integrated into Qwen2.5-7B combined with sparse attention, our BCSR kernel contributes to $2.66{\times}$ end-to-end prefill speedup over dense inference at 64K-token context length with 90\% FFN block sparsity.
\end{itemize}

The remainder of this paper is organized as follows.
Section~\ref{sec:background} provides background on the Hopper architecture, related work, and sparse formats.
Section~\ref{sec:method} describes our kernel designs and the performance analysis of each hardware-specific optimization.
Section~\ref{sec:evaluation} presents our experimental evaluation.
Section~\ref{sec:related} discusses limitations and future directions, and Section~\ref{sec:conclusion} concludes.


%% file: Sec2-Background.tex
\section{Background and Related Work}\label{sec:background}

\subsection{The NVIDIA Hopper Architecture}\label{sec:bg-hopper}

\begin{figure}[t]
\centering
\includegraphics[width=0.85\columnwidth]{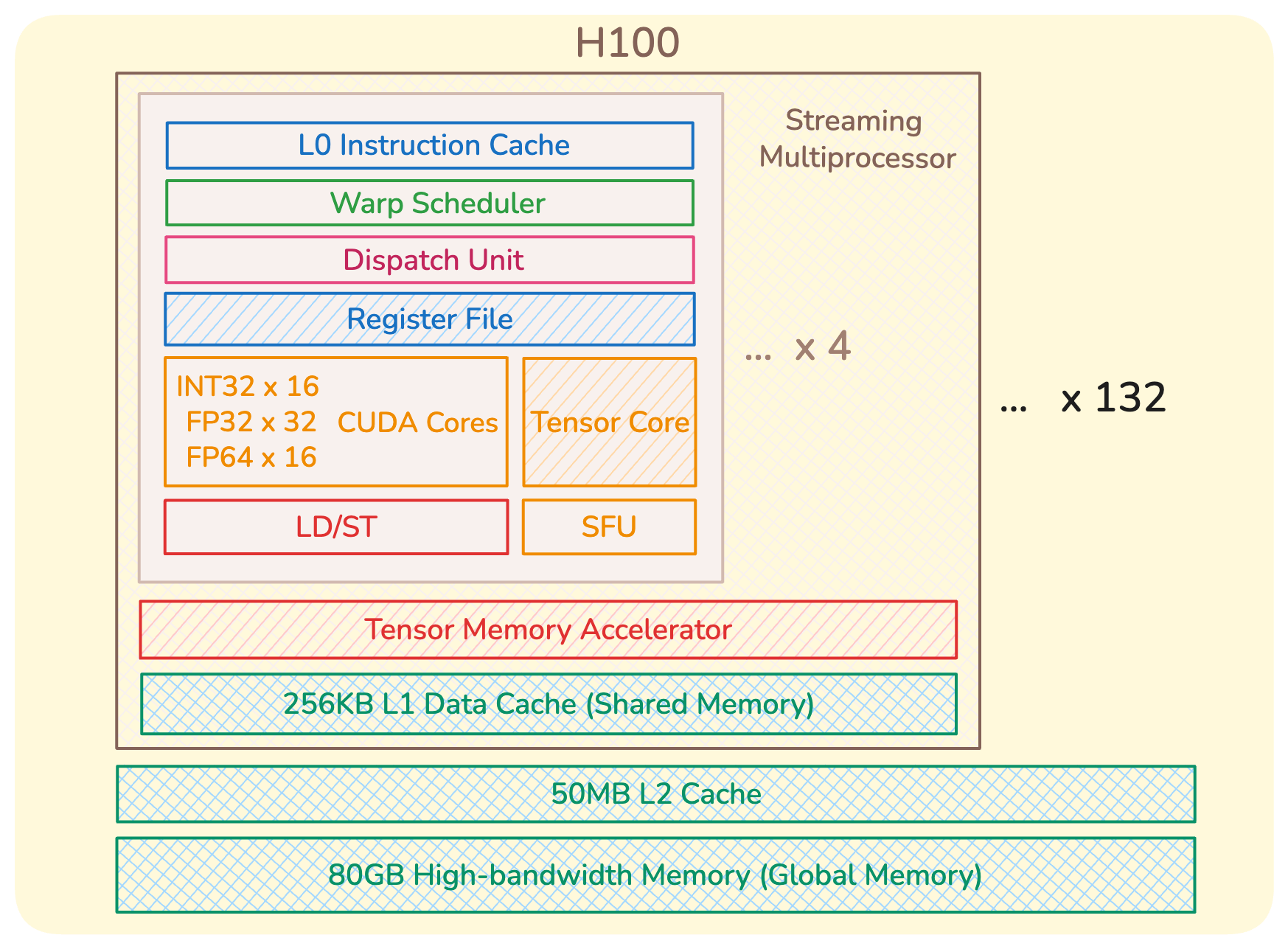}
\caption{An overview of the NVIDIA Hopper (SM90) GPU architecture. Each of 132 SMs contains a TMA unit, a 256\,KB shared memory, and 4 processing blocks with CUDA cores and a 4th generation Tensor Core. All SMs share a 50\,MB L2 cache backed by 80\,GB high-bandwidth global memory (HBM3).}
\label{fig:hopper}
\end{figure}

The NVIDIA H100 (Hopper, SM90) represents a fundamental shift from the synchronous, thread-cooperative programming model of prior architectures (Ampere, Ada Lovelace) to an asynchronous, role-specialized model that decouples data movement from computation~\cite{luo2025dissecting, cudaforfun_h100}.
The H100 architecture is illustrated in Figure~\ref{fig:hopper}. We rely on the following newly introduced architectural features.


\textbf{Tensor Cores and WGMMA.}
On Ampere and Ada, each warp (32~threads) issues a synchronous \code{mma.sync} instruction that reads both operands from registers and writes results back to registers.
Hopper introduces a warpgroup-level (128~threads) \code{wgmma.mma_async} instruction that differs from the prior \code{mma.sync} in three fundamental aspects. First, \code{wgmma} is issued collectively by a warpgroup of 128 threads (four contiguous warps). Second, \code{wgmma} sources one or both operands directly from shared memory, reducing register pressure. Third, \code{wgmma} executes asynchronously on the Tensor Core Unit (TCU) and is ordered through an explicit \code{wgmma.fence} / \code{wgmma.commit_group} / \code{wgmma.wait_group} protocol, whereas \code{mma.sync} is synchronous. These differences allow \code{wgmma} to compute much larger tiles while freeing CUDA cores for other work.
Microbenchmarks show that \code{wgmma} achieves over 96\% of Hopper's 989~TFLOP/s BF16 peak when $N \geq 64$, whereas the legacy \code{mma.sync} path reaches only ${\sim}$63\% of peak~\cite{luo2025dissecting}. To avoid bank conflicts, operands in shared-memory must be laid out with a swizzled pattern, which can be applied automatically by Tensor Memory Accelerator (TMA).



\textbf{Tensor Memory Accelerator (TMA).} On pre-Hopper GPUs, loading a tile into shared memory requires all threads to cooperate: each computes a source address, issues a global load, and writes to shared memory. Hopper's TMA replaces this with a dedicated hardware engine: a single thread issues \code{cp.async.bulk.tensor} with tile coordinates, and TMA handles all address computation, swizzle, and transfers up to 5D tensors between global and shared memory. A host-side tensor map descriptor encodes shape, strides, tile dimensions, and swizzle mode; completion is signaled via barriers. This frees all other threads for useful computation and eliminates 30--40 registers per thread previously consumed by address arithmetic. Its throughput saturates the H100's 3.35~TB/s HBM3 bandwidth when the tile size exceeds 4~KB~\cite{luo2025dissecting}.

\textbf{Warp specialization.}
Because TMA and WGMMA use separate hardware units, peak throughput requires overlapping them in a producer--consumer pipeline. Producer warpgroups issue TMA loads; consumer warpgroups execute WGMMA; synchronization is mediated by per-stage \code{mbarrier} objects in a multi-stage shared-memory circular buffer~\cite{cudaforfun_h100}.
This warp specialization programming paradigm is unavailable on pre-Hopper architectures where all warps must participate in both loading and computing.

\textbf{Thread block clusters and TMA multicast.}
A cluster of up to 16 co-scheduled thread blocks (CTAs) can access each other's shared memory via Distributed Shared Memory (DSMEM)~\cite{luo2025dissecting}.
TMA multicast broadcasts a tile to all cluster members in one transaction, reducing L2 traffic when multiple blocks share the same data (e.g., the same block of $A$ reused across $N$-tiles).

\subsection{Related Work}

\textbf{SpMM Kernels on GPUs.}
Prior to the introduction of TCUs, there are research accelerating SpMM using CUDA cores. For example, GE-SpMM~\cite{gespmm} optimizes SpMM with coalesced row caching and coarse-grained warp merging. Sputnik~\cite{gale2020sparse} accelerate SpMM and SDDMM for sparse matrices in deep learning applications using hierarchical tiling and row-swizzle load balancing.

The emergence of TCUs motivates kernels to exploit dense matrix-multiply hardware for sparse workloads. cuSPARSE~\cite{naumov2010cusparse} serves as the vendor baseline and introduces a TCU-friendly Blocked-ELLPACK format.
TC-GNN~\cite{tcgnn} proposes a CUDA core and TCU collaboration design to accelerate graph neural networks.
DTC-SpMM~\cite{fan2024dtc} combines an efficient sparse format with reordering and runtime optimizations.
FastSpMM~\cite{wang2025fastspmm} remaps long row windows to address load imbalance.
FlashSparse~\cite{shi2025flashsparse} computes transposed sparse matrix multiplication for finer granularity compression.
Acc-SpMM~\cite{zhao2025acc} introduces data-affinity reordering with adaptive load balancing.
However, prior work lacks discussions on leveraging asynchronous GPU architecture features for sparse kernels; this work provides the first systematic study.

Orthogonally, several works preprocess sparse matrices to conform to the N:M structured sparsity pattern required by NVIDIA's sparse tensor cores. Chen et al.~\cite{chen2025accelerating} reorder graph vertices; MP-SpMM~\cite{dong2025bridging} uses matching and padding; TASDER~\cite{jeong2025enabling} decomposes matrices into sums of structured sparse matrices; and Jigsaw~\cite{jigsaw} applies multi-granularity reordering.
N:M preprocessing is beyond the scope of this paper but represents an interesting extension.

\textbf{Tensor Algebra Compilers.} General sparse tensor algebra compilers, such as TACO~\cite{taco}, SparseTIR~\cite{sparsetir}, and UniSparse~\cite{unisparse} offers competitive performance compared to kernels in vendor libraries.
At application level, compiler frameworks such as SPLAT~\cite{gupta2025splat} optimizes regular attention sparsity patterns using a domain-specific sparse format.
Modern GPU compilers such as Triton~\cite{triton} and Tawa~\cite{chen2025tawaautomaticwarpspecialization} support asynchronous features for NVIDIA GPUs such as TMA and warp specialization, but they only support dense tensors.

\textbf{Sparsity in Efficient Machine Learning.} Network pruning has long been utilized to enhance computational efficiency with minimal accuracy degradation in traditional deep learning~\cite{lecun1989optimal, hassibi1992second, han2015learning, sparsevit}. In the era of Large Language Models (LLMs), sparsity has become a critical mechanism on two fronts: compressing model weights~\cite{frantar2023sparsegpt, sun2023simple, xia2023flash, ashkboos2024slicegpt} and extending long-context capabilities via sparse attention mechanisms~\cite{nawrot2026sparsefrontiersparseattention, xiao2024efficient, beltagy2020longformer}. Across both domains, enforcing block-sparse patterns has emerged as a dominant strategy to maximize hardware utilization and memory bandwidth~\cite{xu2025xattentionblocksparseattention,zhu2025efficient,ilin2025thanosblockwisepruningalgorithm,MInference}.
\subsection{Sparse Formats}\label{sec:bg-bsr}

\begin{figure}[t]
\centering
\includegraphics[width=0.95\columnwidth]{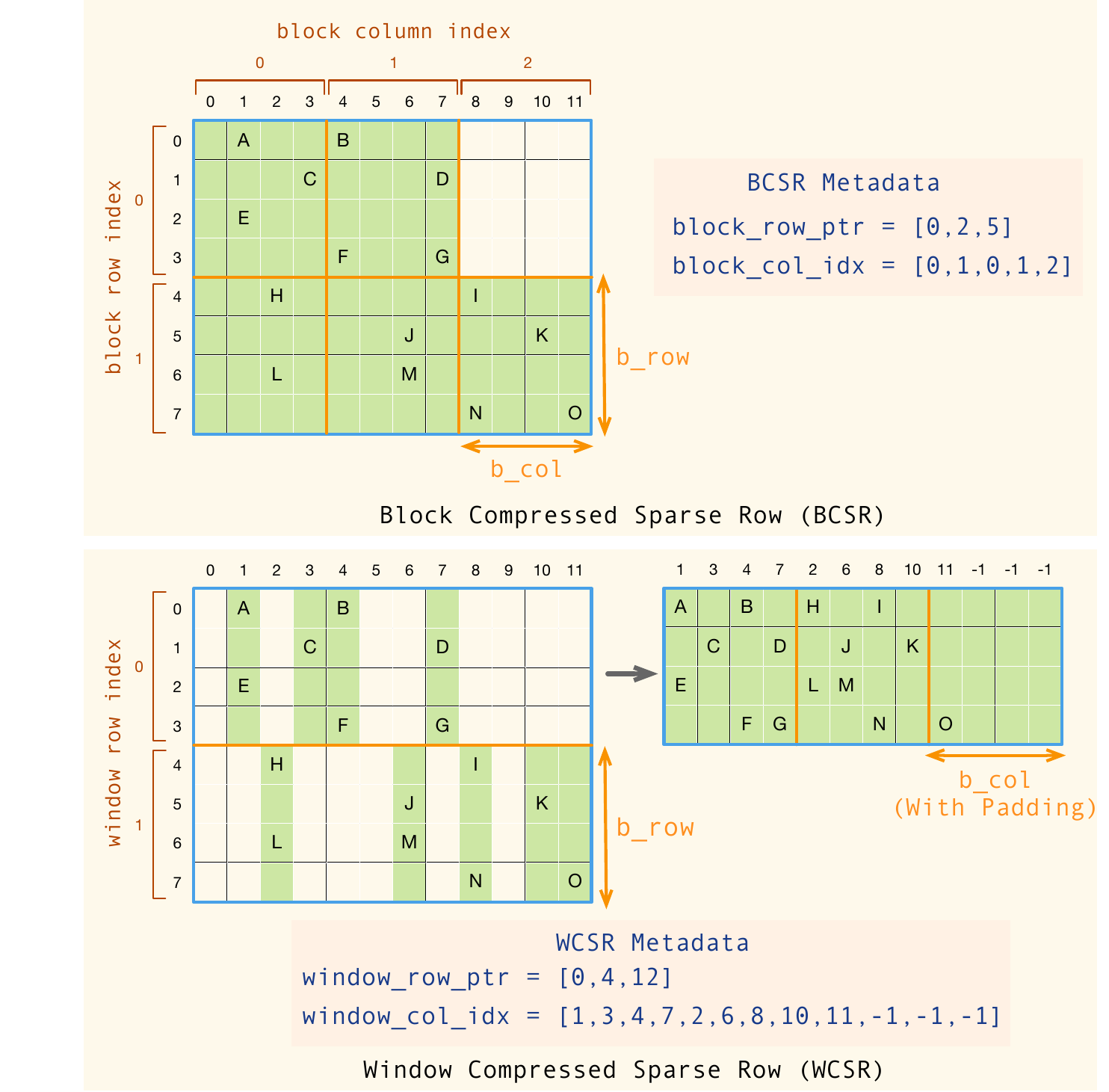}
\caption{Comparison of BCSR and WCSR sparse formats for the same sparse matrix. BCSR tiles both dimensions into fixed $b\_row{\times}b\_col$ blocks, while WCSR groups rows into windows of fixed height ($b\_row$, to be consistent with BCSR), collects non-zero column vectors per window and padded to $b\_col$.}
\label{fig:bcsr-wcsr}
\end{figure}


We consider multiplying a sparse matrix $A \in \mathbb{R}^{m\times k}$ with a dense matrix $B \in \mathbb{R}^{k\times n}$ to produce $C = A \times B$.
We use two sparse formats illustrated in Figure~\ref{fig:bcsr-wcsr} as vehicles to demonstrate the performance impact of hardware features for SpMM kernels.

\textbf{Block Compressed Sparse Row (BCSR).}
BCSR~\cite{im1998model} tiles $A$ into a grid of $b\_row\times b\_col$ blocks and stores only blocks that contain at least one nonzero element.
As shown in Figure~\ref{fig:bcsr-wcsr}, any block that has nonzeros is retained as a dense $b\_row\times b\_col$ tile, while all-zero blocks are discarded.
Three arrays encode the structure: \code{block_row_ptr}$[\frac{m}{b\_row}+1]$ gives the starting index of nonzero blocks in each block-row, \code{block_col_idx}$[\mathit{nnz\_blocks}]$ records the block-column index for each nonzero block, and \code{blocks}$[\mathit{nnz\_blocks} \times b\_row \times b\_col]$ stores the dense values.
The value storage cost is $O(\mathit{nnz\_blocks}\cdot b\_row \cdot b\_col)$ and the index overhead is $O(\frac{m}{b\_row}+\mathit{nnz\_blocks})$, making BCSR efficient for matrices where nonzeros naturally cluster into dense blocks.
Because each nonzero block occupies a contiguous $b\_row \times b\_col$ region in memory, TMA can load it in a single 2D bulk transfer with hardware-applied swizzle.
However, within each nonzero block, zero-valued entries are stored and computed, wasting both storage and arithmetic.
We refer to the fraction of actual nonzeros within stored blocks as the fill ratio: $\mathit{fill\_ratio} = \mathit{nnz} / (\mathit{nnz\_blocks} \cdot b\_row \cdot b\_col)$.
Matrices with scattered nonzeros exhibit low fill ratios and high storage overhead under BCSR.

\textbf{Window Compressed Sparse Row (WCSR).}
FlashSparse~\cite{shi2025flashsparse} compresses the sparse operand using 8$\times$1 column vectors by grouping every 8 rows into a window and storing only non-zero columns present in that window.
In this paper, we build on this idea and generalize it by parameterizing the window size. We call the resulting format Window-Compressed Sparse Row (WCSR).

WCSR uses a finer-grained compression that avoids the rigid $b\_row\times b\_col$ block structure. 
As shown in Figure~\ref{fig:bcsr-wcsr}, rows are grouped into fixed-height windows of $b\_row$ rows, and within each window, it stores the union of all column indices that appear in any of the $b\_row$ rows.
Values are packed into a dense 2D array of shape $[b\_row, \frac{nnz\_cols}{b\_col}+1]$, where each window's columns are padded to a multiple of $b\_col$.
Three arrays encode the structure: \code{window_row_ptr}$[\frac{m}{b\_row}+1]$ gives the starting index of column vectors for each window, \code{window_col_idx}$[\mathit{padded\_nnz\_cols}]$ records the original column index for each packed column position (-1 for padded colum vectors), and \code{values}$[b\_row \times \mathit{padded\_nnz\_cols}]$ stores the packed values.
Unlike BCSR, where both the sparse $A$ values and the dense $B$ rows can be loaded continuously from memory, WCSR requires an indirect address translation to fetch $B$.


\textbf{Trade-offs.}
The two formats exhibit complementary strengths.
BCSR's fixed block structure enables TMA-friendly contiguous access for both $A$ and $B$ operands, but wastes storage on zeros within blocks for matrices with scattered sparsity patterns.
WCSR also incurs padding overhead when the number of non-zero columns is not a multiple of $b\_col$, but is much more compact than BCSR by storing only non-zero columns per window. However, WCSR requires indirect $B$ access via \code{window_col_idx}, introducing data loading overhead.

%% file: Sec3-Method.tex
\section{AsyncSparse}\label{sec:method}


In this section, we incrementally study the impact of each asynchronous GPU architecture feature on our AsyncSparse SpMM kernel performance. 
\subsection{Asynchronous Tensor Core Execution via WGMMA}\label{sec:wgmma}


H100 offers wgmma instructions for shapes \code{m64n{8, 16, 24, ..., 256}k16} on BF16, and \code{m64n{8, 16, 24, ..., 256}k8} on TF32. 
Both our BCSR and WCSR formats choose $b\_row$ to be 64 that matches \code{m}$=64$ for wgmma instructions. The grid is shaped as $(\frac{M}{64},\, \frac{N_{\text{dense}}}{\mathit{BN}})$, where each thread block processes one block-row of $A$ against a $\mathit{BN}$-wide column slice of $B$. 

To add this feature independently, we use cooperative threads to load the non-zero blocks of $A$ and the corresponding tiles of $B$ into shared memory with a swizzled mode to avoid bank conflicts. After a \code{__syncthreads()} barrier, the warpgroup issues consecutive instructions (e.g., $\frac{64}{16} = 4$ \code{wgmma.mma_async.m64n{BN}k16.f32.bf16.bf16} for BF16 kernel, one per $K{=}16$ slice), bracketed by \code{wgmma.fence} / \code{commit_group} / \code{wait_group} to order asynchronous Tensor Core operations. A second \code{__syncthreads()} ensures all WGMMA reads from shared memory complete before the next iteration overwrites the same address with data loading. The temporary accumulation results of $C$ are stored in registers. 

This synchronous load--compute pattern already delivers a substantial speedup over the scalar baseline, as shown in the experimental results of our ablation study (Section~\ref{sec:ablation}, design \textbf{opt1} v.s. \textbf{opt0}). The tile size $\mathit{BN}$ is a free parameter, and its performance impact is analyzed in Section~\ref{sec:eval-tile}. The bottleneck of using WGMMA alone remains the global-memory latency. Tensor Cores sit idle during loads, and the memory pipeline is unused during WGMMA.


\subsection{Asynchronous Data Movement via TMA}\label{sec:tma}

\begin{figure}[t]
\centering
\includegraphics[width=\columnwidth]{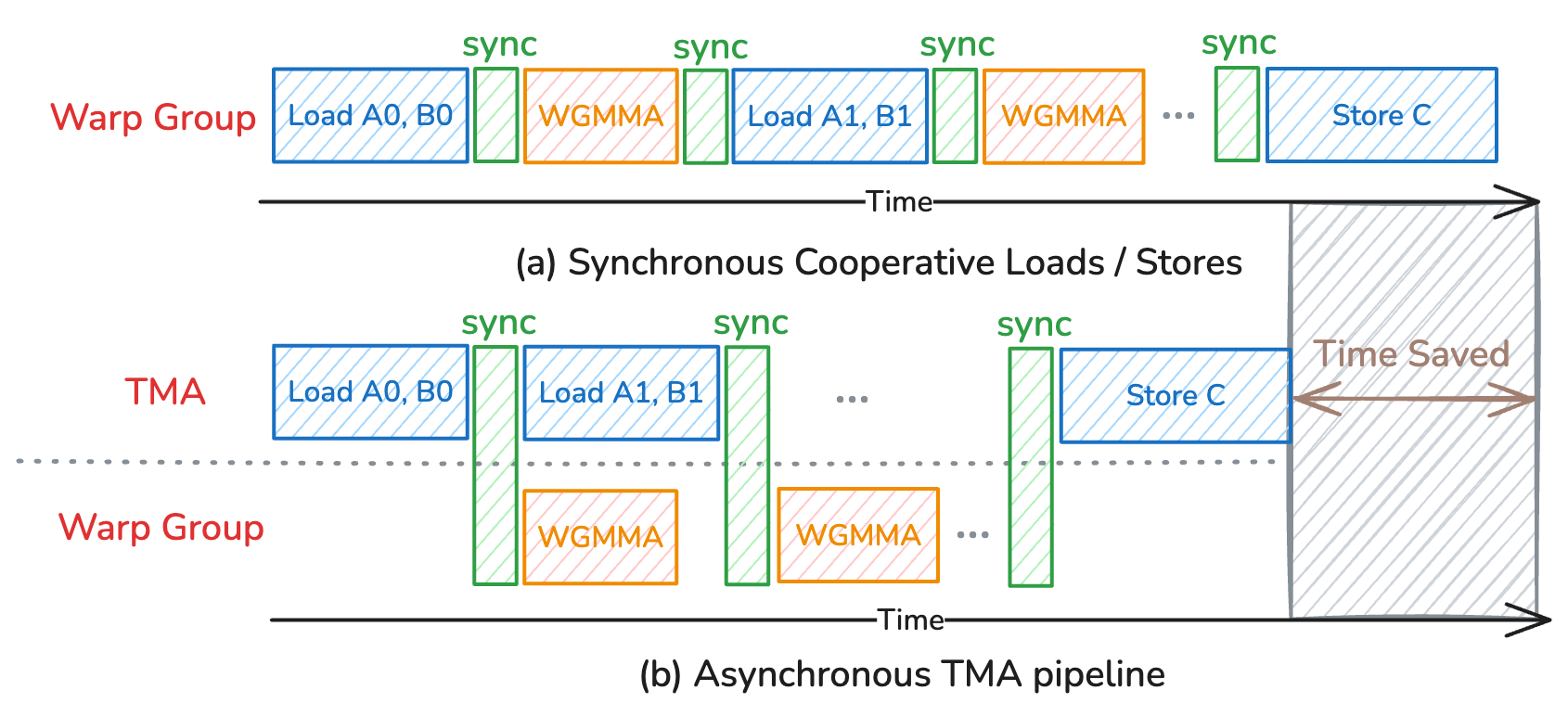}
\caption{Execution timeline.
(a)~Synchronous cooperative loads: all threads alternately load and compute, with synchronization barriers between each phase.
(b)~Asynchronous TMA pipeline: the TMA engine and Tensor Cores operate concurrently on separate hardware.}
\label{fig:pipeline}
\end{figure}

Using the cooperative loading model incurs several drawbacks: address arithmetic consumes registers and issue slots across every thread; shared-memory bank conflicts must be manually avoided through software swizzle patterns; and the load and compute phases are strictly serialized, as illustrated in Figure~\ref{fig:pipeline}a.

Hopper's Tensor Memory Accelerator (TMA) is a dedicated hardware engine that eliminates these costs~\cite{luo2025dissecting}. A single thread issues a \code{cp.async.bulk.tensor} instruction specifying only tile coordinates; the TMA unit handles all address computation, performs the data transfer with hardware-applied swizzle, and signals completion through memory barriers. At runtime, the instruction is non-blocking: \code{cp.async.bulk.tensor} returns immediately, and the transfer proceeds on dedicated hardware, freeing the issuing thread and all other threads to perform useful computation. 
For implementation, we use a host-side descriptor \code{CUtensorMap} created via \code{cuTensorMapEncodeTiled}, which encodes the tensor's data type, dimensions, strides, tile (box) shape, and swizzle mode.

We apply TMA to both our BCSR and WCSR SpMM kernels. In BCSR, each nonzero block stores a dense $64 \times 64$ tile of values at a contiguous address, and the corresponding $64 \times \mathit{BN}$ tile of $B$ is likewise contiguous in the column-major layout. We use one thread to issue two TMA loads per nonzero block -- one for the $A$ block (indexed by the BCSR \code{col_idx}) and one for the $B$ tile, and all other threads remain uninvolved. After the $K$-reduction loop, the accumulated results are stored back to global memory via a single-thread TMA bulk store (\code{cp.async.bulk.tensor.2d.global.shared::cta.tile.bulk_group}), avoiding a cooperative global store by all 128 threads.

In WCSR, the packed values of $A$ within each window remain contiguous and can be loaded via TMA in the same manner. However, the corresponding rows of $B$ are not contiguous: the \code{window_col_idx} array maps each packed column position to randomly accessed rows of $B$, requiring an indirect data loading that TMA cannot perform. Therefore, we fall back to use all threads in the warpgroup cooperatively gather $B$ values where each thread reads a \code{window_col_idx} entry, fetch the source row of $B$ from global memory, and stores it into shared memory with a manually applied 128-byte swizzle pattern. This cooperative $B$-loading keeps all threads occupied during the load phase, in contrast to the BCSR kernel where only one thread is required.

The non-blocking nature of TMA enables multi-stage software pipelining (Figure~\ref{fig:pipeline}b). By allocating a circular buffer of $Q$ stages in shared memory, TMA loads for stage $i{+}1$ can proceed concurrently with WGMMA computation on stage~$i$, overlapping global-memory latency with Tensor Core execution. The ablation study (Section~\ref{sec:ablation}, \textbf{opt2} v.s. \textbf{opt1}) quantifies the throughput improvement by enabling TMA for data transfer.




\subsection{Warp Specialization}\label{sec:warp-spec}

The TMA pipeline decouples data movement from computation, but a single warpgroup that both issues loads and executes WGMMA must still serialize the two phases within its instruction stream. Warp specialization eliminates this serialization by assigning different roles to different warpgroups within the same thread block: a producer warpgroup issues data loads while the remaining consumer warpgroups execute WGMMA. Because TMA and Tensor Cores occupy separate hardware units, the producer and consumers run concurrently when the pipeline is fully warmed up.

\begin{figure}[t]
\centering
\includegraphics[width=\columnwidth]{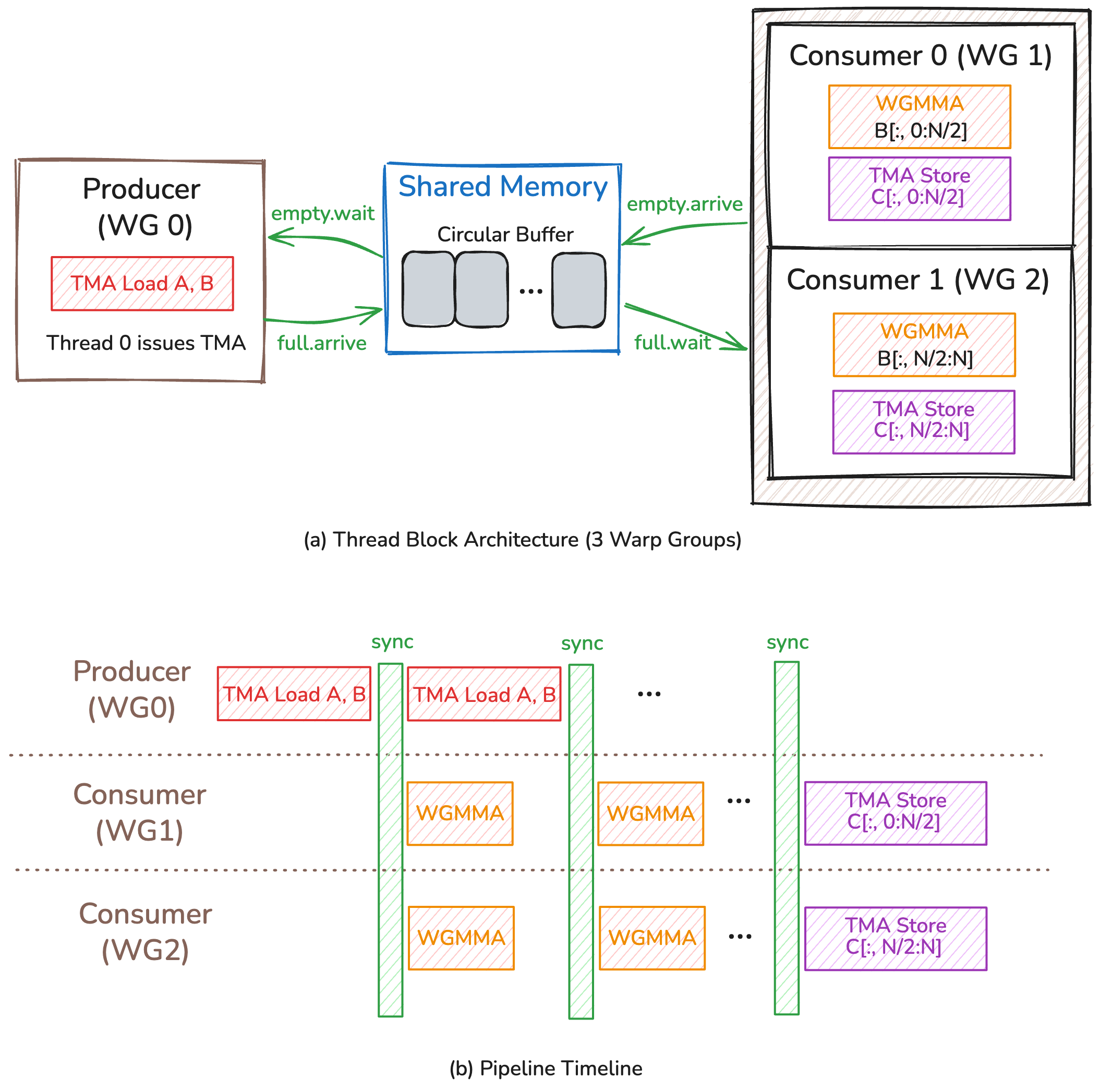}
\caption{Warp-specialization pipeline for the BCSR kernel.
(a)~Thread block architecture: the producer warpgroup (WG\,0) issues TMA loads into a circular buffer in shared memory; two consumer warpgroups (WG\,1--2) execute WGMMA on their respective $N/2$ column slices of $B$ and store disjoint portions of $C$.
(b)~Pipeline timeline of warp-specialization.
The producer overlaps TMA loading with computations in two consumers.}
\label{fig:warp-spec}
\end{figure}

Our BCSR kernel allocates three warpgroups per thread block (384~threads), as shown in Figure~\ref{fig:warp-spec}a. Warpgroup~0 serves as the producer, while warpgroups~1 and~2 serve as consumers. Because TMA handles both $A$ and $B$ loads, only a single thread (thread~0) in the producer warpgroup is active during the load phase; the remaining 127 threads are idle. We use two consumer warp groups to scale up designs using more registers with large $BN$. The two consumers partition the $N$ dimension equally, reading the same $A$-block tile but different column slices of $B$, thereby doubling the compute performed per loaded $A$ tile. After the $K$-reduction loop, each consumer writes its output tile to global memory via TMA bulk store; since the two consumers write to disjoint column ranges, no inter-consumer synchronization is needed.

The producer and consumers communicate through a circular buffer of $Q{=}3$ stages in shared memory, where each stage holds one $64 \times 64$ tile of $A$ and one $64 \times \mathit{BN}$ tile of $B$. Two arrays of shared memory barriers (\code{full[Q]} and \code{empty[Q]}) mediate access using phase-bit tracking. The producer waits on \code{empty[q]}, sets the expected transaction byte count via \code{mbarrier.arrive.expect_tx}, and issues TMA loads whose completion automatically signals \code{full[q]}. Each consumer waits on \code{full[q]}, executes WGMMA, and signals \code{empty[q]} upon finishing. At startup, consumers pre-signal all \code{empty} slots so the producer can begin immediately. The depth $Q{=}3$ ensures that up to two TMA loads are in flight while one stage is being consumed, which suffices to hide the H100's global memory latency (Figure~\ref{fig:warp-spec}b). The pipeline drains naturally when the producer exhausts the nonzero blocks in the current block-row. The ablation performance analysis (Section~\ref{sec:ablation}, \textbf{opt3}) demonstrates a big performance gain with warp specialization.

We use Hopper's \code{setmaxnreg} PTX instruction to enable dynamic register redistribution within a CTA. A per-CTA register pool allows warps to release or acquire registers at runtime: the producer warpgroup executes \code{setmaxnreg.dec.sync.aligned 24} to release registers to the pool, while each consumer warpgroup executes \code{setmaxnreg.inc.sync.aligned 240} to draw from the pool and hold its WGMMA accumulator fragment. Without this reallocation, the combined register demand of three warpgroups would reduce occupancy to one CTA per SM. The same \code{setmaxnreg} instruction must be executed by all warps in a warpgroup, and explicit synchronization is required between successive register adjustments.

The WCSR kernel presents a fundamentally different situation. As described in Section~\ref{sec:tma}, loading $B$ in WCSR requires all threads in the producer warpgroup to cooperatively gather values via indirect \code{window_col_idx} lookups, which means the producer warpgroup is fully occupied during the load phase rather than running a single-thread TMA request. This eliminates the key benefit of warp specialization: the producer is no longer free to run ahead of the consumers, and dedicating an entire warpgroup to loading provides diminishing returns when all its threads are already busy. Our WCSR kernel therefore uses a single warpgroup of 128~threads that performs both loading and computation in each iteration. Thread~0 issues a TMA load for the contiguous $A$ tile while all 128 threads cooperatively gather $B$ values into shared memory. After a synchronization barrier and a TMA completion wait, the same warpgroup executes WGMMA collectively. To address the load imbalance inherent in WCSR where windows vary widely in column count, the kernel employs a task-based decomposition: large windows are split into fixed-size sub-tasks, and \code{blockIdx.x} maps to task descriptors rather than windows directly. When multiple tasks process the same window, output correctness is ensured via \code{atomicAdd} on the output matrix.



\subsection{Thread Block Clusters and TMA Multicast}\label{sec:cluster}

When the grid partitions the $N$ dimension across multiple thread blocks, adjacent blocks along $N$ that belong to the same block-row of $A$ all require the same $A$ tile but different column slices of $B$. Each block independently loading the same $A$ tile wastes L2 cache bandwidth. Hopper introduces thread block clusters, which co-schedule a configurable group of CTAs onto adjacent SMs and enable direct access to each other's shared memory. Combined with TMA multicast, a single \code{cp.async.bulk.tensor} instruction with the \code{.multicast::cluster} qualifier and a bitmask selecting the destination CTAs can broadcast the $A$ tile to all cluster members' shared memories in one L2 transaction, eliminating redundant loads.

Our BCSR kernel sets the cluster size to \code{CLUSTER_N}${=}2$ along the $N$ dimension. Within each cluster, the rank-0 CTA's producer issues the TMA multicast load for the shared $A$ tile, while every CTA's producer issues its own unicast TMA load for its respective $B$ tile. This halves the L2 read traffic for $A$ compared to independent loading. The barrier protocol adapts to the cross-CTA scope: the \code{empty} barrier is initialized with an arrival count of \code{num_consumers}$\times$\code{CLUSTER_N}, and consumers signal the barrier of remote CTAs via the \code{mapa.shared::cluster} address mapping instruction followed by \code{mbarrier.arrive.shared::cluster}. Similarly, consumers wait on the \code{full} barrier with \code{.acquire.cluster} scope to ensure visibility of the multicast data.

The ablation performance analysis (Section~\ref{sec:ablation}, \textbf{opt7}) reveals a throughput regression when enabling 2-CTA multicast, which is primarily attributed to additional scheduling and synchronization overhead.


\subsection{L2 Cache Optimizations}\label{sec:l2}

In dense GEMM, persistent kernels unblock two L2 cache optimizations~\cite{cudaforfun_h100}. The first is overlapping stores of the current tile with TMA loads of the next tile within the same persistent block. The second is program id (PID) swizzling, a column-major remapping of thread block indices that ensures simultaneously active SMs process spatially adjacent output tiles, maximizing reuse of $B$-tile data in the L2 cache. Specifically, output tiles are grouped into vertical strips of \code{GROUP_M} consecutive $M$-tiles, and within each strip the tiles are visited in column-major order so that tiles sharing the same $N$-coordinate are processed on neighbor SMs at the same time, increasing the probability that their shared $B$ columns remain in L2.

We implemented both for our BCSR SpMM kernel but find they regress performance (Section~\ref{sec:ablation}, \textbf{opt6}) for two reasons.
First, PID swizzling assumes $M$-tiles at the same $N$-coordinate share $B$ loads, but in SpMM each $M$-tile loads $B$ rows indexed by its own BCSR \code{block_col_idx}, which rarely overlap across row blocks, leading to limited reuse benefit.
Second, the persistent kernel's static round-robin assignment creates severe load imbalance. SMs that are assigned dense block rows become bottleneck while those with sparse rows idle, negating inter-tile overlap benefits.

\subsection{Load Balance}\label{sec:load-balance}

The load-imbalance problem motivates dynamic tile scheduling.
We implemented two approaches: an atomic tile counter where each producer atomically increments a global counter to claim tiles, and a work-stealing variant using the PTX mbarrier API with embedded tile metadata.

Despite achieving better load balance, neither outperforms the static non-persistent kernel. First, the global atomicAdd serializes all tiles' stores; when tiles have few nonzero blocks, producers reclaim tiles at high frequency, making the atomic a throughput bottleneck.
Second, dynamic scheduling introduces per-tile metadata communication overhead which is not present in the static design, where both producer and consumer derive tile assignments from \code{blockIdx}.
Third, persistent kernels must re-zero accumulators for every new tile, which is well amortized in dense GEMM (hundreds of iterations per tile) but poorly amortized in SpMM, where many tiles have much fewer nonzero blocks.

The non-persistent kernel (Section~\ref{sec:warp-spec}) therefore remains fastest.
The CUDA runtime's built-in scheduler provides dynamic load balancing -- blocks assigned to sparse rows complete quickly and their SM resources are reclaimed for pending dense-row blocks, without atomic overhead or accumulator resets.

%% file: Sec4-Evaluation.tex
\section{Evaluation}\label{sec:evaluation}


All experiments are conducted on a single NVIDIA H100 GPU with 96\,GB of HBM3 memory (3.35\,TB/s peak bandwidth) and 132 streaming multiprocessors clocked at up to 1{,}980\,MHz.
Kernels are compiled with CUDA Toolkit 12.6 (\code{nvcc} 12.6.85) and GCC 11.5.0 as the host compiler, targeting compute capability \code{sm_90a}.
Each kernel is first executed for 10 warm-up iterations to stabilize GPU clock frequencies and caches; execution time is then measured over 100 timed iterations using \code{cudaEvent} timestamps bracketing the kernel launch.
To ensure a fair comparison across formats with different block sizes and padding overheads, we report throughput (TFLOP/s) as $(2 \times \mathit{nnz} \times N) / t$, where $\mathit{nnz}$ is the number of nonzeros in the original sparse matrix, $N$ is the dense-matrix width, and $t$ is the measured kernel time.

\subsection{SpMM Evaluation}\label{sec:eval-performance}
\textbf{Baselines and datasets.}
In this section, we compare our implementations against six SpMM baselines: AccSpMM~\cite{zhao2025acc}, FlashSparse~\cite{shi2025flashsparse}, DTC-SpMM~\cite{fan2024dtc}, TC-GNN~\cite{tcgnn}, TorchAO~\cite{or2025torchao}, and cuSPARSE Blocked-Ellpack (BELL) SpMM (v12.5.8).
We construct our BCSR format with block size (\code{b_row} and \code{b_col}) 64, and our WCSR format with window siz (\code{b_row}) 64 and padding size (\code{b_col}) 8. Each matrix is preprocessed with Reverse Cuthill Mckee Algorithm~\cite{rcm} implemented in \code{scipy} library to help improve non-zero locality.
All comparisons use TF32 precision, as AccSpMM, DTC-SpMM, and TC-GNN only support TF32. We evaluate on 414 sparse matrices from the SuiteSparse Matrix Collection~\cite{davis2011university}, following the same selection as DTC-SpMM~\cite{fan2024dtc}.

Table~\ref{tab:spmm-performance} reports geomean TFLOPS and WCSR speedup over each baseline, stratified by matrix density and dense-matrix width~$N$. Our WCSR kernel achieves the highest geomean throughput in every configuration, reaching 23.53~TFLOPS at $N{=}1024$ on matrices with density ${\geq}1\%$, a $4.86{\times}$ speedup over cuSPARSE and $2.40{\times}$ over FlashSparse. Our BCSR kernel underperforms on the full dataset (``All''), as the $64{\times}64$ blocking introduces substantial zero-padding overhead (low fill ratio) for highly sparse matrices. However, BCSR surpasses all baselines at density ${\geq}0.1\%$ and narrows the gap with WCSR at higher densities, reaching 18.05~TFLOPS ($1.30{\times}$ behind WCSR) at density ${\geq}1\%$, $N{=}1024$.


Another notable trend is that the baselines (AccSpMM, FlashSparse, DTC-SpMM) see their advantage over cuSPARSE erode as matrix density grows. For instance, looking at the TFLOPS numbers, AccSpMM's speedup drops from $2.81/0.72 \approx 5.58{\times}$ on all matrices to just $8.27/5.01 \approx 1.68{\times}$ at density ${\geq}1\%$ for ($N{=}1024$). In contrast, our WCSR maintains a $4$--$6{\times}$ speedup over cuSPARSE across all density thresholds. TC-GNN and torchao are consistently slower than cuSPARSE on denser matrices (density~${\geq}0.5\%$), with TC-GNN dropping to $2.12/3.87 \approx 0.55{\times}$ and torchao to $3.09/3.87 \approx 0.80{\times}$ for density ${\geq}0.5\%$ at $N{=}1024$.


Our kernels scale well with increasing~$N$: WCSR throughput grows from 17.09 to 20.67~TFLOPS as $N$ increases from 256 to 1024 on matrices with density~${\geq}0.5\%$, and BCSR similarly improves from 11.86 to 14.74~TFLOPS. AccSpMM, in contrast, degrades from 9.38 to 8.13~TFLOPS over the same range, suggesting its scatter--gather data movement becomes bandwidth-limited at wider~$N$. DTC-SpMM cannot run at $N{=}1024$ because its kernel allocates $N/16$ warps per thread block, exceeding the hardware limit of 1024~threads.

\begin{figure*}[t]
\centering
\includegraphics[width=\textwidth]{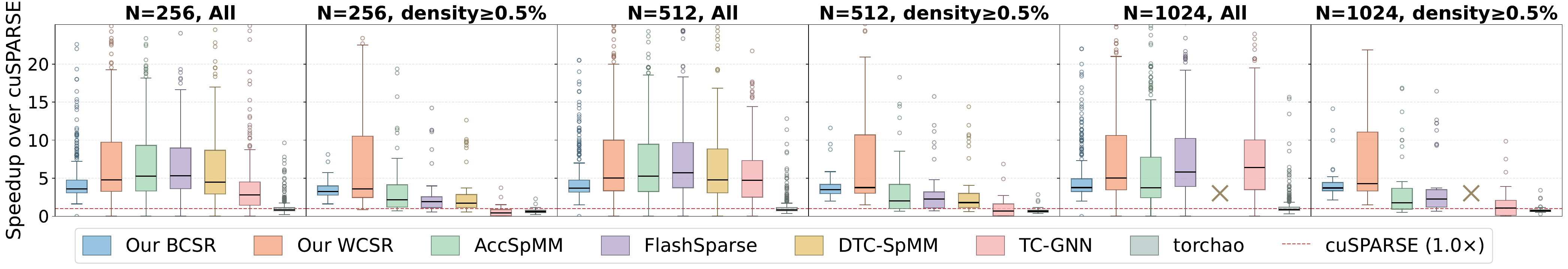}
\caption{Distribution of normalized speedup over cuSPARSE BELL SpMM on SuiteSparse matrices for $N=256$, $512$, and $1024$. The dashed red line marks $1\times$ cuSPARSE baseline.}
\label{fig:spmm-boxplot}
\end{figure*}

The box plots in Figure~\ref{fig:spmm-boxplot} show the per-matrix speedup distribution over cuSPARSE. Our WCSR kernel shows a slightly lower median speedup compared with FlashSparse at $N=512$ and $N=1024$, because FlashSparse's $8{\times}1$ window incurs less column-union padding than our $64{\times}1$ window on the most extremely sparse matrices (density~$<0.1\%$). However, on matrices with density~${\geq}0.5\%$, both WCSR and BCSR dominate all baselines in median, confirming that the asynchronous features (TMA+WGMMA) are  particularly effective once the sparse format overhead is amortized. TorchAO remains below cuSPARSE across all settings, indicating that Triton's generic code generation does not exploit TMA or warp specialization effectively for sparse workloads.


\begin{table*}[t]
\centering
\caption{Geomean TFLOPS on SuiteSparse matrices at different density thresholds. Each cell shows TFLOPS (WCSR speedup over this kernel).}
\label{tab:spmm-performance}
\normalsize
\renewcommand{\arraystretch}{1.0}
\resizebox{\textwidth}{!}{
\begin{tabular}{l|cccc|cccc|cccc}
\toprule
\multirow{2}{*}{\textbf{Kernel}} & \multicolumn{4}{c}{\textbf{$N=256$}} & \multicolumn{4}{c}{\textbf{$N=512$}} & \multicolumn{4}{c}{\textbf{$N=1024$}} \\
\cmidrule(lr){2-5}
\cmidrule(lr){6-9}
\cmidrule(lr){10-13}
 & All & $\geq$0.1\% & $\geq$0.5\% & $\geq$1\% & All & $\geq$0.1\% & $\geq$0.5\% & $\geq$1\% & All & $\geq$0.1\% & $\geq$0.5\% & $\geq$1\% \\
\midrule
\textbf{Our WCSR} & \textbf{3.86} & \textbf{10.59} & \textbf{17.09} & \textbf{18.92} & \textbf{4.04} & \textbf{11.76} & \textbf{19.33} & \textbf{21.89} & \textbf{4.13} & \textbf{12.38} & \textbf{20.67} & \textbf{23.53} \\
Our BSR & \makecell{2.45 (1.58$\times$)} & \makecell{7.54 (1.41$\times$)} & \makecell{11.86 (1.44$\times$)} & \makecell{14.04 (1.35$\times$)} & \makecell{2.56 (1.58$\times$)} & \makecell{8.48 (1.39$\times$)} & \makecell{13.61 (1.42$\times$)} & \makecell{16.45 (1.33$\times$)} & \makecell{2.62 (1.58$\times$)} & \makecell{8.97 (1.38$\times$)} & \makecell{14.74 (1.40$\times$)} & \makecell{18.05 (1.30$\times$)} \\
AccSpMM & \makecell{3.57 (1.08$\times$)} & \makecell{7.32 (1.45$\times$)} & \makecell{9.38 (1.82$\times$)} & \makecell{9.46 (2.00$\times$)} & \makecell{3.54 (1.14$\times$)} & \makecell{7.30 (1.61$\times$)} & \makecell{9.16 (2.11$\times$)} & \makecell{9.35 (2.34$\times$)} & \makecell{2.81 (1.47$\times$)} & \makecell{6.29 (1.97$\times$)} & \makecell{8.13 (2.54$\times$)} & \makecell{8.27 (2.85$\times$)} \\
FlashSparse & \makecell{3.46 (1.12$\times$)} & \makecell{6.06 (1.75$\times$)} & \makecell{7.47 (2.29$\times$)} & \makecell{7.84 (2.41$\times$)} & \makecell{3.78 (1.07$\times$)} & \makecell{7.09 (1.66$\times$)} & \makecell{8.57 (2.26$\times$)} & \makecell{8.94 (2.45$\times$)} & \makecell{3.91 (1.05$\times$)} & \makecell{7.78 (1.59$\times$)} & \makecell{9.52 (2.17$\times$)} & \makecell{9.79 (2.40$\times$)} \\
DTC-SpMM & \makecell{3.21 (1.20$\times$)} & \makecell{5.91 (1.79$\times$)} & \makecell{7.68 (2.23$\times$)} & \makecell{8.31 (2.28$\times$)} & \makecell{3.39 (1.19$\times$)} & \makecell{6.64 (1.77$\times$)} & \makecell{8.40 (2.30$\times$)} & \makecell{8.91 (2.46$\times$)} & --- & --- & --- & --- \\
TC-GNN & \makecell{1.17 (3.32$\times$)} & \makecell{0.83 (12.88$\times$)} & \makecell{0.71 (24.25$\times$)} & \makecell{0.58 (32.90$\times$)} & \makecell{1.98 (2.04$\times$)} & \makecell{1.47 (8.11$\times$)} & \makecell{1.26 (15.27$\times$)} & \makecell{1.04 (20.93$\times$)} & \makecell{2.79 (1.47$\times$)} & \makecell{2.30 (5.48$\times$)} & \makecell{2.12 (9.79$\times$)} & \makecell{1.84 (12.94$\times$)} \\
torchao & \makecell{0.66 (6.17$\times$)} & \makecell{1.49 (7.12$\times$)} & \makecell{2.12 (8.19$\times$)} & \makecell{2.38 (8.16$\times$)} & \makecell{0.70 (6.06$\times$)} & \makecell{1.74 (6.76$\times$)} & \makecell{2.64 (7.44$\times$)} & \makecell{3.20 (7.05$\times$)} & \makecell{0.71 (6.11$\times$)} & \makecell{1.90 (6.52$\times$)} & \makecell{3.09 (6.81$\times$)} & \makecell{3.91 (6.23$\times$)} \\
cuSPARSE & \makecell{0.72 (5.82$\times$)} & \makecell{1.94 (5.47$\times$)} & \makecell{3.68 (4.71$\times$)} & \makecell{4.67 (4.16$\times$)} & \makecell{0.72 (6.04$\times$)} & \makecell{2.02 (5.84$\times$)} & \makecell{3.83 (5.13$\times$)} & \makecell{4.96 (4.55$\times$)} & \makecell{0.72 (6.24$\times$)} & \makecell{2.01 (6.18$\times$)} & \makecell{3.87 (5.45$\times$)} & \makecell{5.01 (4.86$\times$)} \\
\bottomrule
\end{tabular}
}
\renewcommand{\arraystretch}{1.0}
\end{table*}

\input{Sec4-2-ablation}
\input{Sec4-3-tile-size}
\input{Sec4-4-llm}

%% file: Sec4-2-ablation.tex
\subsection{Ablation Study}\label{sec:ablation}
\textbf{Baselines and datasets.}
We isolate the performance contribution of each async hardware feature using the BF16 BCSR kernel, an important data type for machine learning workloads. We compare against cuSPARSE BELL and TorchAO's Triton-based BSR kernel as the only SpMM baselines supporting BF16, and evaluate on the same 414 SuiteSparse matrices as in Section~\ref{sec:eval-performance} with $N=1024$.

\begin{figure}[t]
\centering
\includegraphics[width=0.87\columnwidth]{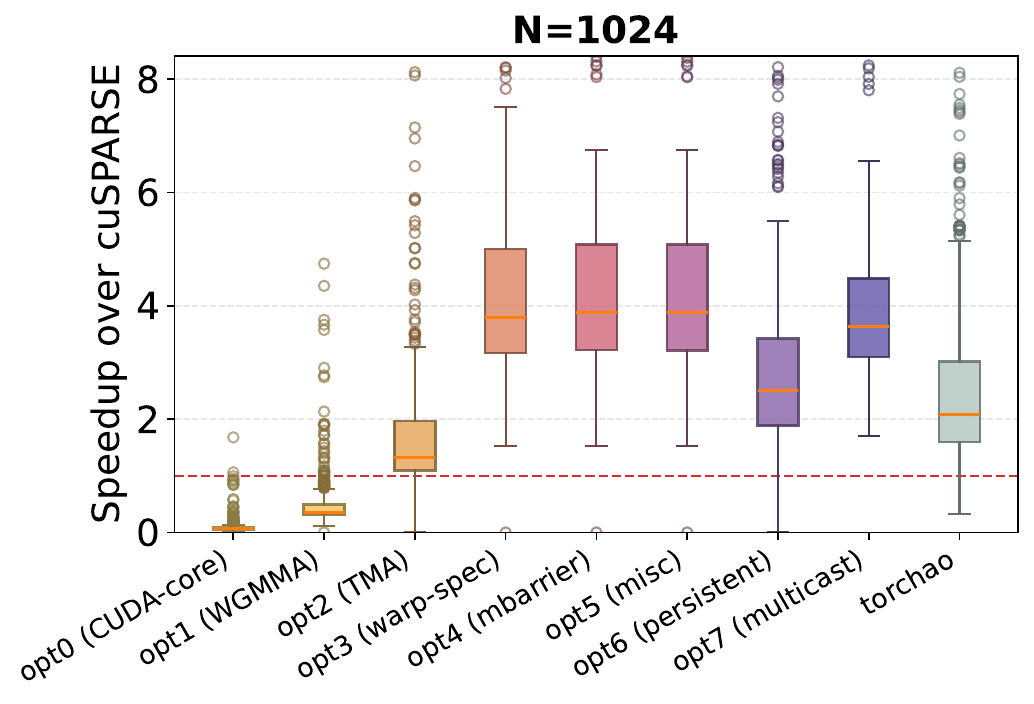}
\caption{Performance breakdown showing the cumulative impact of each optimization on SuiteSparse matrices with $N=1024$, normalized to cuSPARSE BELL SpMM (dashed line at $1\times$).}
\label{fig:perf-comparison}
\end{figure}

Figure~\ref{fig:perf-comparison} shows the per-matrix speedup distribution at each progressive optimization stage, normalized to cuSPARSE BELL SpMM (dashed line at $1\times$).
Table~\ref{tab:bf16-ablation} reports the corresponding geomean TFLOPS and speedup. We incrementally enable optimizations across eight stages, from a na\"ive CUDA-core baseline to more complicated designs.

\textbf{opt0: CUDA-core baseline ($\boldsymbol{0.08\times}$).}
A thread-cooperative BSR SpMM kernel using 128 threads and scalar FMA instructions on CUDA cores. It achieves 0.12 geomean TFLOPS ($0.08\times$ cuSPARSE BELL), confirming that scalar execution is far from competitive for SpMM workloads on Hopper.

\textbf{opt1: opt0 + WGMMA ($\boldsymbol{+0.34\times}$).}
Replacing scalar FMAs with \code{wgmma.mma_async} Tensor Core instructions raises throughput to 0.63~TFLOPS ($0.42\times$), a $5.3\times$ improvement over opt0. The kernel employs 128B swizzled shared memory layout and \code{fence.proxy.async} to ensure correct data visibility across WGMMA's async proxy domain. Despite the large compute gain, synchronous global-memory loads leave the Tensor Core pipeline underutilized, keeping throughput below cuSPARSE.

\textbf{opt2: opt1 + TMA ($\boldsymbol{+1.14\times}$).}
Replacing thread-cooperative loads with TMA \code{cp.async.bulk} operations yields 2.37~TFLOPS ($1.56\times$), the first configuration that outperforms cuSPARSE. TMA offloads address generation and data movement to dedicated hardware, freeing warps for compute. Its asynchronous nature enables overlapping the next tile's data transfer with the current tile's WGMMA execution.


\textbf{opt3: opt2 + Warp specialization ($\boldsymbol{+2.75\times}$).}
Dedicating one warp group as producer (TMA loads) and two warp groups as consumers (WGMMA compute), connected by a three-stage circular buffer, raises throughput to 6.44~TFLOPS ($4.31\times$). This achieves the single largest incremental gain, contributing $+2.75\times$ cuSPARSE, nearly two-thirds of the total improvement from opt0 to this point. The producer--consumer pipeline eliminates bubbles: compute warps execute WGMMA on the current buffer slot while the producer simultaneously issues TMA loads for the next.

\textbf{opt4: opt3 + raw mbarrier ($\boldsymbol{+0.10\times}$).}
Replacing C++ \code{cuda::barrier} with raw PTX \code{mbarrier} instructions and manual phase-bit tracking raises throughput to 6.59~TFLOPS ($4.41\times$). This reduces synchronization overhead by having only one thread per warp group arrive at each barrier, rather than all 128 threads.

\textbf{opt5: opt4 + accumulator zero-elision ($\boldsymbol{\approx 0\times}$).}
Eliminating the explicit accumulator \code{memset} by using \code{ScaleD=0} on the first WGMMA iteration maintains throughput at 6.58~TFLOPS ($4.40\times$). This micro-optimization has negligible impact at the geomean level, indicating the kernel is approaching the memory-bandwidth ceiling after warp specialization has saturated the compute pipeline.

The following two configurations branch independently from opt5 to explore whether the benefits of persistent kernels and TMA multicast in dense GEMM translate to the sparse setting

\textbf{opt6: opt5 + persistent kernel ($\boldsymbol{-1.68\times}$).}
Switching to a persistent kernel with PID swizzling reduces throughput to 4.08~TFLOPS ($2.72\times$). This design launches a fixed grid of ${\#SM}$ thread blocks that loop over output tiles in a static order. The persistent loop introduces load imbalance issues for sparse workloads.
Unlike dense GEMM, sparse workloads exhibit highly variable per-tile computation: some output tiles map to many nonzero blocks while others map to few.
This imbalance leaves thread blocks idle after finishing sparse tiles, negating the launch-overhead and L2-locality benefits of persistence.

\textbf{opt7: opt5 + TMA multicast ($\boldsymbol{-0.30\times}$).}
Adding thread-block clusters (\code{__cluster_dims__(1, CLUSTER_N, 1)}) with TMA multicast and write-through streaming stores (\code{__stwt()}) also regresses from opt5.
TMA multicast shares A-tile loads across 2 CTAs in a cluster, reducing redundant global memory traffic, while each CTA independently loads its own B tile. However, clusters impose a co-scheduling constraint: all \code{CLUSTER_N} CTAs must launch simultaneously on adjacent SMs, reducing the scheduler's flexibility to balance irregular sparse workloads. Furthermore, the cross-CTA \code{mbarrier} synchronization, where consumers across all CTAs in the cluster must signal before the producer can reuse a buffer slot, introduces higher latency than CTA-local barriers. For diverse sparse matrices where per-tile work varies widely, these overheads outweigh the bandwidth savings from multicast.



\textbf{Summary.}
WGMMA ($+0.34\times$), TMA ($+1.14\times$), and warp specialization ($+2.75\times$) account for $+4.31\times$ out of the $+4.41\times$, approximately $98\%$ improvement from opt0 to opt4.
Persistent kernels and TMA multicast show benefit for dense GEMM kernels, but they regress in the sparse setting, which underscores that hardware features for dense linear algebra do not automatically benefit sparse workloads.

\begin{table}[t]
\centering
\caption{BF16 kernel ablation on 414 SuiteSparse matrices ($N{=}1024$).
  Each cell shows geomean TFLOPS (speedup over cuSPARSE).}
\label{tab:bf16-ablation}
\small
\resizebox{0.89\columnwidth}{!}{
\begin{tabular}{lc}
\toprule
\textbf{Kernel} & \textbf{TFLOPS (speedup)} \\
\midrule
opt0 (CUDA-core) & 0.12 (0.08$\times$) \\
opt1 (WGMMA) & 0.63 (0.42$\times$) \\
opt2 (TMA) & 2.37 (1.56$\times$) \\
opt3 (warp-spec) & 6.44 (4.31$\times$) \\
opt4 (mbarrier) & 6.59 (4.41$\times$) \\
opt5 (misc) & 6.58 (4.40$\times$) \\
opt6 (persistent) & 4.08 (2.72$\times$) \\
opt7 (multicast) & 10.05$^{*}$ (4.10$\times$) \\
TorchAO & 4.00 (2.42$\times$) \\
cuSPARSE BELL & 1.60 \\
\bottomrule
\multicolumn{2}{p{0.9\columnwidth}}{\scriptsize $^{*}$Over 300 matrices; 114 fail cluster launch. cuSPARSE geomean on this subset is 2.45 TFLOPS.}
\end{tabular}
}
\end{table}

%% file: Sec4-3-tile-size.tex
\subsection{Tile Size Selection}\label{sec:eval-tile}
The WGMMA instruction fixes the tile height at \code{m=}64 and the reduction dimension at \code{k=}16 for BF16, leaving the tile width $\mathit{WGMMA\_N}$ in \code{m64n{WGMMA_N}k16} as a free parameter that can be chosen from 8 to 256 in steps of 8. In our warp-specialized kernel, two consumer warpgroups each process $\mathit{WGMMA\_N}$ columns, so the total tile width is $\mathit{BN}{=}2\times \mathit{WGMMA\_N}$ and ranges from 16 to 512. This parameter significantly affects performance because it controls both the amount of useful computation performed per loaded $A$ tile and the SM resource consumption that determines occupancy.

\begin{figure}[t]
\centering
\includegraphics[width=\columnwidth]{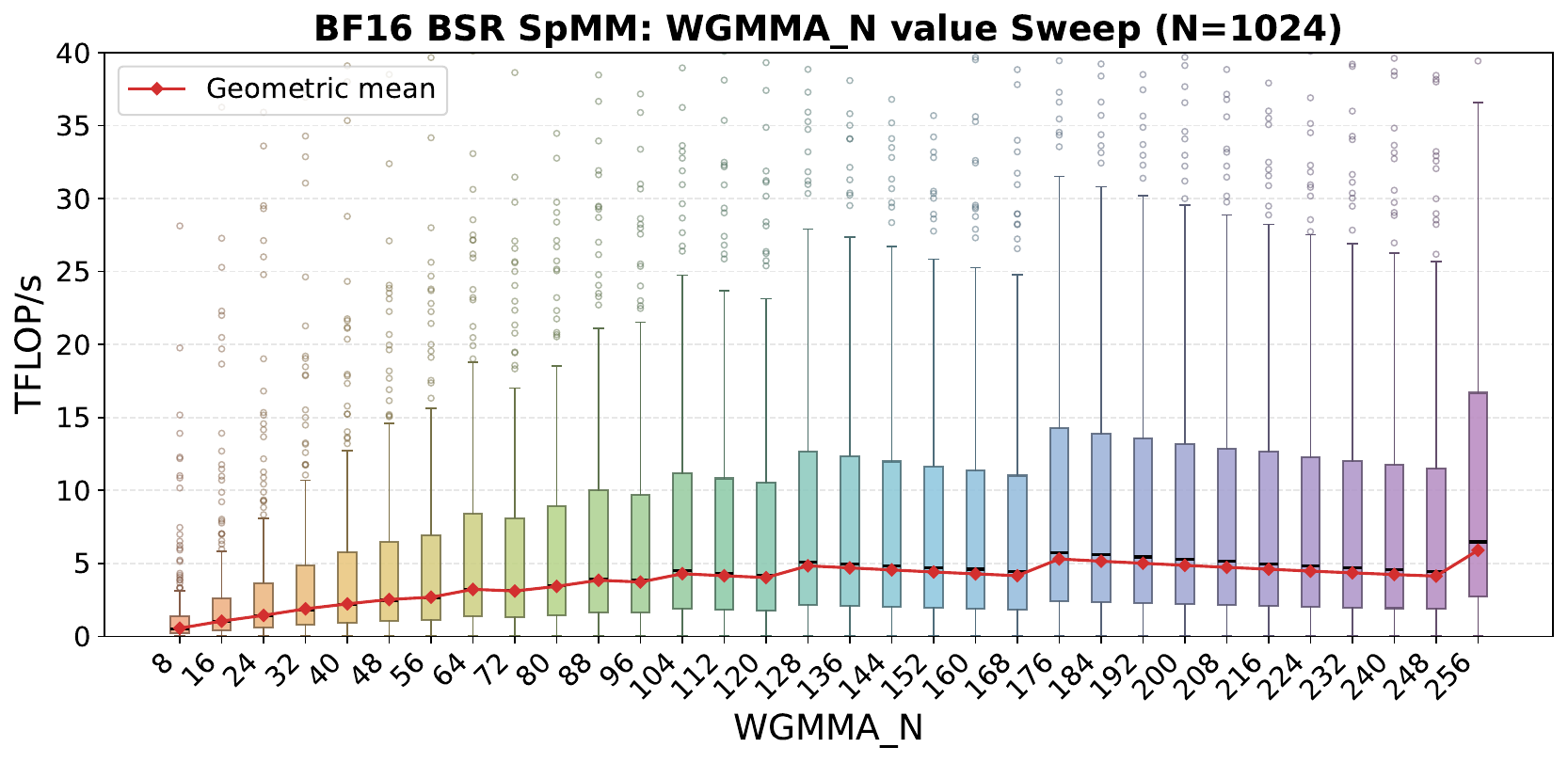}
\caption{Warp-specialized kernel throughput variance with $\mathit{WGMMA\_N}$ from 8 to 256 (step 8) across SuiteSparse matrices with $\mathit{N}{=}1024$.}
\label{fig:sweep-n}
\end{figure}

Figure~\ref{fig:sweep-n} shows the throughput distribution across the 414 SuiteSparse matrices (same selection as \ref{sec:eval-performance} and \ref{sec:ablation}) as $\mathit{WGMMA\_N}$ varies from 8 to 256 with dense matrix width $\mathit{N}{=}1024$. The geometric mean throughput rises from 0.56\,TFLOP/s at $\mathit{WGMMA\_N}{=}8$ to 5.90\,TFLOP/s at $\mathit{WGMMA\_N}{=}256$, a $10.5\times$ improvement that demonstrates larger tiles amortize TMA load overhead and barrier synchronization cost.

We also observe performance drops from tile padding cost. When $\mathit{BN}{=}2\times \mathit{WGMMA\_N}$ does not evenly divide $\mathit{N}$, the kernel must pad the dense matrix to the next multiple of $\mathit{BN}$, computing on zero-filled columns that consume time without contributing useful throughput. $\mathit{WGMMA\_N}$ values whose $\mathit{BN}$ divides 1024, such as $\mathit{WGMMA\_N}{=}256$, 128, and 64, incur no padding and form local peaks. $\mathit{WGMMA\_N}{=}176$ ($\mathit{BN}{=}352$) also peaks because its padding overhead of only 3\% is much smaller compared to its neighbors. The penalty grows with $\mathit{WGMMA\_N}$ because each additional padded tile represents a larger fraction of total work. For example, $\mathit{WGMMA\_N}{=}248$ ($\mathit{BN}{=}496$) pads $\mathit{N}$ from 1024 to 1488, wasting 45\% of the computation and reducing its geometric mean to 4.13\,TFLOP/s, below that of the much smaller $\mathit{WGMMA\_N}{=}176$ at 5.30\,TFLOP/s.

Based on this analysis, we select the max $\mathit{WGMMA\_N}$ size that is divisible by the input $\mathit{N}$ for our kernels, as it avoids padding waste and achieves the highest geometric mean throughput per our evaluation.

%% file: Sec4-4-llm.tex
\subsection{Case Study: End-to-End LLM Prefill}\label{sec:eval-e2e-llm}
We integrate our BCSR kernel into the PyTorch framework to evaluate its efficacy in speeding up the prefill phase of large language models (LLMs). Transformer layer computation is primarily bottlenecked by two operations: self-attention, whose cost scales quadratically with sequence length, and the feed-forward network (FFN), whose cost scales linearly.
We sparsify the LLM under various configurations to explore the attainable speedups, deliberately relaxing accuracy constraints to focus on the upper bound of performance gains.
Our BCSR SpMM kernel targets accelerating the three large matrix multiplications in the FFN projections. We use MInference~\cite{MInference} to accelerate self-attention which profiles heads offline to identify dominant block-sparse patterns and dynamically applies the best-fitting pattern at inference time. We integrate both our BCSR kernel for the FFN and MInference for self-attention into Qwen2.5-7B~\cite{qwen2.5} to evaluate the overall end-to-end prefill latency across various sequence lengths.

\textbf{Model and integration.} Qwen2.5-7B is a 28-layer transformer with a hidden dimension $h{=}3{,}584$ and a SwiGLU intermediate dimension $d{=}18{,}944$. This yields three FFN projections per layer (\code{gate_proj} and \code{up_proj} mapping $h \to d$; \code{down_proj} mapping $d \to h$), all of which have dimensions divisible by our $64 \times 64$ block size. We apply random block sparsity at 90\% and 95\% to the FFN weights. We then replace each sparsified \code{nn.Linear} with a custom drop-in module that invokes our BCSR kernel directly. MInference patches the self-attention forward pass to compute only the dynamically selected sparse subset of query--key blocks using its Triton-based fused attention kernel.

\textbf{Kernel-level FFN speedup.}
Each FFN projection computes $C = W_{\text{sparse}} \times X^{\top}$, where the sparse weight $W$ has shape $M \times K$ and the dense input $X$ has shape $N \times K$ with $N$ equal to the sequence length. We isolate the \code{gate_proj} projection ($M{=}18{,}944$, $K{=}3{,}584$) in Qwen2.5-7B and compare our BCSR kernel on sparsified weights against dense \code{torch.mm} that calls cuBLAS. We measure the \code{gate_proj} latency in BF16 precision, averaged over 100 runs after 10 warm-up iterations.
Table~\ref{tab:qwen-kernel} reports the results across four sparsity levels and four sequence lengths. At 90\% block sparsity, our kernel achieves $1.58\times$ to $1.98\times$ speedup over cuBLAS; at 99\% sparsity the speedup reaches up to $3.19\times$.
The speedup scales consistently with sparsity, confirming that our BCSR kernel's throughput improves with the reduction in non-zero blocks.

\begin{table}[t]
\centering
\caption{Qwen2.5-7B gate\_proj kernel latency comparison between dense torch.mm (torch) and our BCSR. The numbers in parentheses are speedup relative to torch.mm.}
\label{tab:qwen-kernel}
\setlength{\tabcolsep}{2pt}
\footnotesize
\resizebox{\columnwidth}{!}{
\begin{tabular}{c|cc|cc|cc|cc}
\toprule
 & \multicolumn{2}{c|}{80\%} & \multicolumn{2}{c|}{90\%} & \multicolumn{2}{c|}{95\%} & \multicolumn{2}{c}{99\%} \\
\cmidrule(lr){2-3} \cmidrule(lr){4-5} \cmidrule(lr){6-7} \cmidrule(lr){8-9}
$N$ & torch & BCSR & torch & BCSR & torch & BCSR & torch & BCSR \\
\midrule
1{,}024  & 0.24 & 0.17 (1.46$\times$) & 0.22 & 0.14 (1.58$\times$) & 0.23 & 0.12 (1.89$\times$) & 0.23 & 0.09 (2.60$\times$) \\
4{,}096  & 0.86 & 0.55 (1.57$\times$) & 0.87 & 0.44 (1.98$\times$) & 0.85 & 0.39 (2.18$\times$) & 0.85 & 0.28 (3.04$\times$) \\
16{,}384 & 3.27 & 2.06 (1.59$\times$) & 3.27 & 1.70 (1.92$\times$) & 3.27 & 1.46 (2.24$\times$) & 3.33 & 1.05 (3.19$\times$) \\
65{,}536 & 13.57 & 8.12 (1.67$\times$) & 13.12 & 6.69 (1.96$\times$) & 13.12 & 5.88 (2.23$\times$) & 12.87 & 4.21 (3.06$\times$) \\
\bottomrule
\end{tabular}
}
\end{table}

\begin{figure}[t]
\centering
\includegraphics[width=0.9\columnwidth]{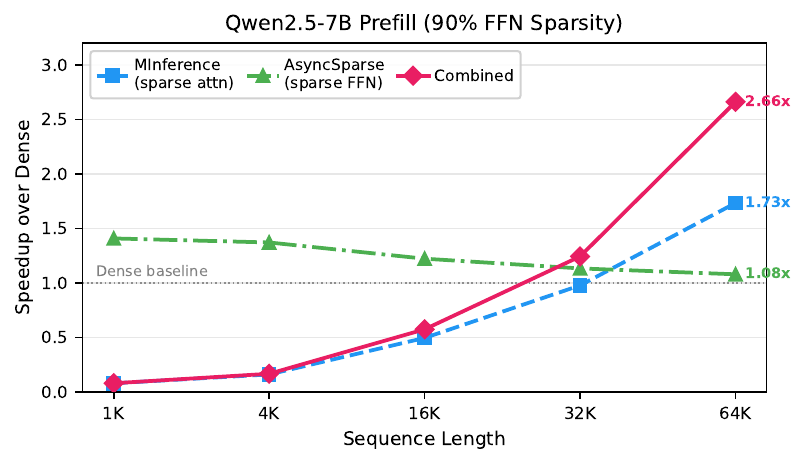}
\caption{End-to-end prefill speedup over the dense baseline on Qwen2.5-7B (90\% FFN block sparsity).
  MInference accelerates attention at long sequences; our BCSR kernel accelerates FFN at all lengths.
  The combined approach exceeds either technique alone, reaching $2.66\times$ at 64K tokens.}
\label{fig:qwen-e2e-speedup}
\end{figure}

\textbf{End-to-end prefill latency.}
We measure the full-model prefill latency under four configurations: dense baseline, MInference alone (sparse attention, dense FFN), our BCSR kernel alone (dense attention, sparse FFN), and the combination of both.
The dense baseline uses PyTorch's SDPA for attention, which dispatches to cuDNN's Hopper-native flash attention kernel, and cuBLAS for all linear projections.
MInference replaces the attention kernel with its own Triton-based sparse kernel.
All configurations use BF16 precision on a single H100 GPU.
Figure~\ref{fig:qwen-e2e-speedup} reports the speedup of each configuration relative to the dense baseline across sequence lengths from 1{,}024 to 65{,}536 at 90\% FFN block sparsity.

At short sequence lengths ($N \leq 4{,}096$), the FFN projections constitute the dominant compute cost because the quadratic attention term remains negligible relative to the three linear FFN projections per layer. Our sparse FFN kernel reduces this bottleneck, delivering $1.37\times$ to $1.41\times$ end-to-end speedup. MInference incurs overhead at short sequences because its Triton-based sparse attention introduces fixed costs such as dynamic pattern selection, and these costs exceed the savings from skipping attention blocks when the attention computation itself is small.

As the sequence length grows, the $O(N^2)$ attention cost overtakes the $O(N)$ FFN cost. At $N{=}32{,}768$, MInference reaches parity with the dense baseline ($0.98\times$), and at $N{=}65{,}536$ it delivers $1.73\times$ speedup by computing only the dynamically selected subset of attention blocks. Our FFN kernel's relative contribution diminishes in this regime because dense attention increasingly dominates the total latency, yielding only $1.08\times$ at $N{=}65{,}536$.

The combined configuration reveals that the two sparsification techniques address orthogonal bottlenecks and compose favorably. At $N{=}65{,}536$, the combination achieves $2.66\times$ speedup, substantially exceeding MInference alone ($1.73\times$) and our FFN kernel alone ($1.08\times$). This effect arises because MInference reduces the attention cost while our kernel reduces the FFN cost, and the two savings stack without interference.

%% file: Sec5-Limitations.tex
\section{Limitations and Future Work}\label{sec:related}
\textbf{Limitations}. Both BCSR and WCSR formats fix the block or window height at 64 rows to match the WGMMA $m{=}64$ tile dimension; support finer granularities could reduce padding overhead and would require redesigning the TMA and WGMMA pipelines. Additionally, this work does not investigate matrix preprocessing algorithms; advanced row and column reordering design may produce more efficient layouts by clustering nonzeros into denser blocks.

\textbf{Future directions}.
FP8 WGMMA support would further increase throughput and directly benefit quantized LLM inference. Multi-GPU scaling via NCCL would enable matrices exceeding single-GPU memory capacity. Load imbalance remains a challenge, and leveraging architectural features in newer GPU generations such as Cluster Launch Control on NVIDIA Blackwell is a new direction.
More broadly, investigating how TMA-based loads/stores and warp-specialized pipelining transfer to other irregular workloads beyond SpMM and to newer generations of GPUs is a natural continuation of this work.

%% file: Sec6-Conclusion.tex
\section{Conclusion}\label{sec:conclusion}
This paper presented AsyncSparse, a study of co-designing high-performance SpMM kernels on asynchronous GPU architectures. We demonstrate the integration of asynchronous GPU programming models for SpMM using two sparse formats.
Our WCSR kernel combines TMA loads with cooperative thread gathering and employs row spliting for load balance, which achieves state-of-the-art performance on SuiteSparse matrices.
Our BCSR kernel leverages a warp-specialized pipeline that overlaps TMA-driven data loads/stores with WGMMA Tensor Core computation, and it contributes to a combined $2.66{\times}$ end-to-end prefill speedup on Qwen2.5-7B at 64K-token context length.
Our ablation study showed that WGMMA, TMA, and warp specialization account for approximately 98\% of the total improvement over a CUDA-core baseline, while persistent kernels and TMA multicast regress in our SpMM implementation due to load imbalance and co-scheduling constraints.